\documentclass
[preprint,byrevtex,10pt,twocolumn,tightenlines,prl,letterpaper,nobalancelastpage,noshowpacs]{revtex4}%
\usepackage{amsfonts}
\usepackage{amsmath}
\usepackage{amssymb}
\usepackage[dvips]{graphicx}
\usepackage{hyphenat}%
\providecommand{\U}[1]{\protect\rule{.1in}{.1in}}

\begin{document}
\preprint{ }
\title{On the relation between disorder and homogeneity in an amorphous metal}
\author{Z. Ovadyahu}
\affiliation{Racah Institute of Physics, The Hebrew University, Jerusalem 9190401, Israel }

\begin{abstract}
Disorder and homogeneity are two concepts that refer to spatial variation of
the system potential. In condensed-matter systems disorder is typically
divided into two types; those with local parameters varying from site to site
(diagonal disorder) and those characterized by random transfer-integral values
(off-diagonal disorder). Amorphous systems in particular exhibit off-diagonal
disorder due to random positions of their constituents. In real systems
diagonal and off-diagonal disorder may be interconnected. The formal depiction
of disorder as local deviations from a common value focuses attention on the
short-range components of the potential-landscape. However, long range
potential fluctuation are quite common in real systems. In this work we seek
to find a correlation between disorder and homogeneity using amorphous
indium-oxide (In$_{\text{x}}$O) films with different carrier-concentrations
and with different degree of disorder. Thermal treatment is used as a means of
fine tuning the system disorder. In this process the resistance of the sample
decreases while its amorphous structure and chemical composition is preserved.
The reduced resistivity affects the Ioffe-Regel parameter k$_{\text{F}}\ell$
that is taken as a relative measure of disorder in a given sample. The
homogeneity of the system was monitored using inelastic light-scattering. This
is based on collecting the Raman signal from micron-size spots across the
sample. The statistics of these low-energy data are compared with the sample
disorder independently estimated from transport measurements. The analysis
establishes that heterogeneity and disorder are correlated.

\end{abstract}
\maketitle

\section{ Introduction}

Disorder plays a major role in the properties of condensed-matter systems
\cite{1,2,3,4,5,6,7,8,9}. The experimental study of disorder-induced phenomena
present a challenge in terms of being able to control, characterize, and
quantify it. Effort in this vein was mainly invested in the field of
electronic transport. In particular, this issue has been a major concern in
studies of the metal-insulator transition (MIT) and the
superconductor-insulator transition (SIT). The system conductivity is
frequently used as an empirical measure of disorder in these studies. The
conductivity of a solid is arguably its most sensitive property and it may be
affected by different means, not all of them may be attributed to disorder. A
change in carrier-concentration for example, would affect the conductivity
while causing only a small change in the disorder.

An aspect of disorder frequently referred to but rarely actually measured is
homogeneity. It should be realized that the two concepts are related; spatial
disorder is tantamount to inhomogeneous. Strong disorder, such as required to
Anderson-localize a system with a relatively high carrier-concentration, is
naturally accompanied by significant spatial variations of the potential. We
shall refer to this type of inhomogeneity as `inherent'. A system may however
be inhomogeneous with little or no short-range potential-fluctuations being
present. This form of `technological' inhomogeneity \cite{10,11} may lead to
non-trivial transport effects in otherwise `clean' systems. In principle, the
two types of inhomogeneities may be told apart; changing the disorder should
show a concomitant change of inhomogeneity if the disorder is inherent.

As a relevant example, we present in this note experiments to study the
relation between disorder and spatial inhomogeneity in amorphous indium-oxide
(In$_{\text{x}}$O) films. Using a purely amorphous system for this project is
natural; the only structurally imposed length-scale is the nearest-neighbor
distance which turns out to be much smaller than any dimension relevant to the
problem. This research focused on the version of the material that is
relatively rich in carrier-concentration. This version of In$_{\text{x}}$O has
been extensively studied and its transport parameters are well known
\cite{12,13}. In particular, near the critical point of the MIT the magnitude
of the disorder may be quantitatively determined from the condition for the
Anderson transition. The heterogeneity at this disorder is compared with
another version of In$_{\text{x}}$O that topologically is nearly identical
while much less disordered. Subjecting both systems to thermal treatment that
modifies their resistivities allows for another way to test the effect of
disorder on system homogeneity. The sample disorder before and after heat
treatment is characterized by the Ioffe-Regel parameter k$_{\text{F}}\ell$.
Low-energy Raman-spectroscopy is employed to probe the system spatial
homogeneity in each case for tracking the change associated with the modified
disorder. The results support the expected notion that heterogeneity is an
inherent property of the disordered system.

\section{Experimental}

\subsection{Samples preparation and measurements techniques}

The In$_{\text{x}}$O films used in this work were e-gun evaporated onto
room-temperature substrates using 99.999\% pure In$_{\text{2}}$O$_{\text{3}}$
sputtering-target. Deposition was carried out at the ambience pressure of
3$\pm$0.5x10$^{\text{-5}}$ to 4$\pm$0.5x10$^{\text{-4}}$ Torr oxygen-pressure
maintained by leaking 99.9\% pure O$_{\text{2}}$ through a needle valve into
the vacuum chamber (base pressure $\simeq$10$^{\text{-6}}$ Torr). Undoped
silicon wafers with $\langle$100$\rangle$ orientation were used as substrates
for electrical and Raman measurements. Carbon-coated copper grids were used
for transmission electron-diffraction. During film deposition, the grids were
anchored to 1mm glass-slides by small indium pellets later removed for
mounting the grids in the microscope. The deposited film on the rest of the
slide was used for monitoring the sample resistance for comparison. X-ray
interferometry was performed on samples deposited on a 2.8mm float-glass.

Rates of deposition in the range 0.3-2.5~\AA /s were used to produce films
with different compositions; The In$_{\text{x}}$O samples had
carrier-concentration \textit{N }that increases with the ratio of
deposition-rate to the oxygen-partial-pressure. For the rates-pressures used
here \textit{N} was in the range 1x10$^{\text{19}}$cm$^{\text{-3}}$ to
9x10$^{\text{20}}$cm$^{\text{-3}}$ as measured by Hall effect at
room-temperature using a Hall-bar control-sample prepared simultaneously for
each sample deposition. Lateral sizes of samples used for transport
measurements was typically 1x2.5mm$^{\text{2}}$ (width x length respectively),
and 1x1cm$^{\text{2}}$ for the Raman scattering experiments. The evaporation
source to substrate distance in the deposition chamber was 45$\pm
$1cm.Therefore, thickness variations across the samples are unlikely to be
related to geometrical consideration in the deposition process.

To afford reasonable resolution for electron-microscopy thickness of the films
used for the electron-microscope work was d=20$\pm$1nm.

Two batches of In$_{\text{x}}$O with carrier-concentrations of \textit{N}%
=(1.5$\pm$1)x10$^{\text{19}}$cm$^{\text{-3}}$, and \textit{N}=(8$\pm
$1)x10$^{\text{20}}$cm$^{\text{-3}}$ were deposited for the Raman scattering
experiments. In the following these will be referred to as "low-\textit{N}"
and "high-\textit{N}" versions of In$_{\text{x}}$O respectively. The
high-\textit{N} version may be also referred to in this work as the
"indium-rich" phase of the compound. These version of In$_{\text{x}}$O are
representative for the substance used in recent electron-glass experiments
with typically \textit{N}$\approx$10$^{\text{19}}$cm$^{\text{-3}}$ and the
In$_{\text{x}}$O version with \textit{N}$\approx$10$^{\text{21}}%
$cm$^{\text{-3}}$ commonly used in superconductor-insulator studies
\cite{14,15,16,17,18,19,20,21,22}. 

The Ioffe-Regel parameter k$_{\text{F}}\ell$=(9$\pi^{\text{4}}/$%
\textit{N})$^{\text{1/3}}\frac{\text{R}_{\text{Q}}}{\rho_{\text{RT}}}$ \ where
R$_{\text{Q}}$=$\hslash/$e$^{\text{2}}$ is the resistance quantum and
$\rho_{\text{RT}}$ is the resistivity at room-temperature, is used here as a
measure of the sample relative disorder. This parameter is monotonic with
disorder, and it is well defined experimentally. It is descriptive of the
static disorder even when its value is smaller than unity where neither
k$_{\text{{\small F}}}$ nor $\ell$ are "good" parameters (which is the case in
all our "as-made" samples). The important caveat that should be borne in mind
in this regard is that when k$_{\text{F}}\ell\ll$1, $\ell$ must not be
interpreted as a semiclassical mean-free-path. At room temperatures, where our
measurements are performed, it is safe to say that the transport length-scale
(whether it is the hopping-length or the mean-free-path) is \textit{much}
smaller than the size of the region probed by the laser beam.

Sample homogeneity on a micron-scale was characterized by the intensity
variations of the Raman signal over the energy range 70-420cm$^{\text{-1}}$.
These were carried for In$_{\text{x}}$O samples with different \textit{N} as
well as for the as-made and the annealed samples where disorder was modified
by thermal-treatment.

Electron diffraction patterns were taken with the Philips Tecnai F20 G2
operating at 200kV. The Raman spectra were taken with a Renishaw inVia Reflex
Spectrometer using a laser beam with a wavelength of 514 nm and edge-filter at
$\approx$70cm$^{\text{-1}}$. These measurements employed beam-spots of 2$\mu$m
and 5$\mu$m. In each case, fifty spectra were taken at different spots across
a rectangle area of the sample for obtaining statistics of the Raman signal
magnitude. The laser intensity was checked to be in the linear response and
low enough to avoid structural changes during exposure \cite{23}. During these
preliminary tests, the \textit{same} spot on the sample was measured several
times. Traces were taken from this spot with different laser intensities. The
spot could be identified in the microscope as it changed color after several
exposures. However, the Raman traces taken from it registered the same
spectrum over the range studied with intensity variations of typically $\pm
$0.5\%. Therefore the variations of signal intensity reported below are mostly
due to inhomogeneities rather than due to time dependence. 

The deposited samples, after being characterized, were heat treated to modify
their disorder at T=360$\pm$5K for 24-48 hours. Fuller details related to heat
treatment of In$_{\text{x}}$O films are described elsewhere \cite{24}.

\section{Results and discussion}

Local Raman-spectra for the high-\textit{N} version In$_{\text{x}}$O film are
shown in Fig. 1 for the as-made, indium-rich sample. As mentioned above, this
version of In$_{\text{x}}$O has been often used for the disorder-driven SIT.
This transition occurs when the Ioffe-Regel parameter k$_{\text{F}}\ell$ of
the system is 0.29-0.32 \cite{25}. Figure 2 shows the respective set of data
for the same sample after it has been heat-treated and its k$_{\text{F}}\ell$
value has increased from 0.1 to 0.42. The data in Fig. 1 and Fig. 2 pertain
therefore to the insulating and superconducting side of the SIT respectively.

Two features stand out in these data: First, the local intensity of the Raman
signal fluctuates considerably from point to point (Fig. 1a and Fig. 1c)
manifesting the spatial heterogeneity of the sample. Secondly, homogeneity of
the sample visibly improves after heat-treatment (Fig. 2a and Fig. 2c). The
sample heterogeneity parameter $\gamma$ is defined here as a ratio:
$\gamma\equiv$[standard-deviation]/[mean] of a normal-distribution fitted to
the intensity-histogram. The use of normal distribution to fit the histogram
data is just to allow an impartial estimate of the sample heterogeneity. To
determine the precise functional form of the distribution would require many
more data plots than used here. Nonetheless, the obtained data are consistent
and systematic enough to test the basic conjecture concerning the
disorder-homogeneity relation. Note that $\gamma$ is smaller for the annealed
sample, and it is smaller for the larger spot-size as might be expected from
the better ensemble averaging.%
\begin{figure}[ptb]%
\centering
\includegraphics[
height=2.4448in,
width=3.4411in
]%
{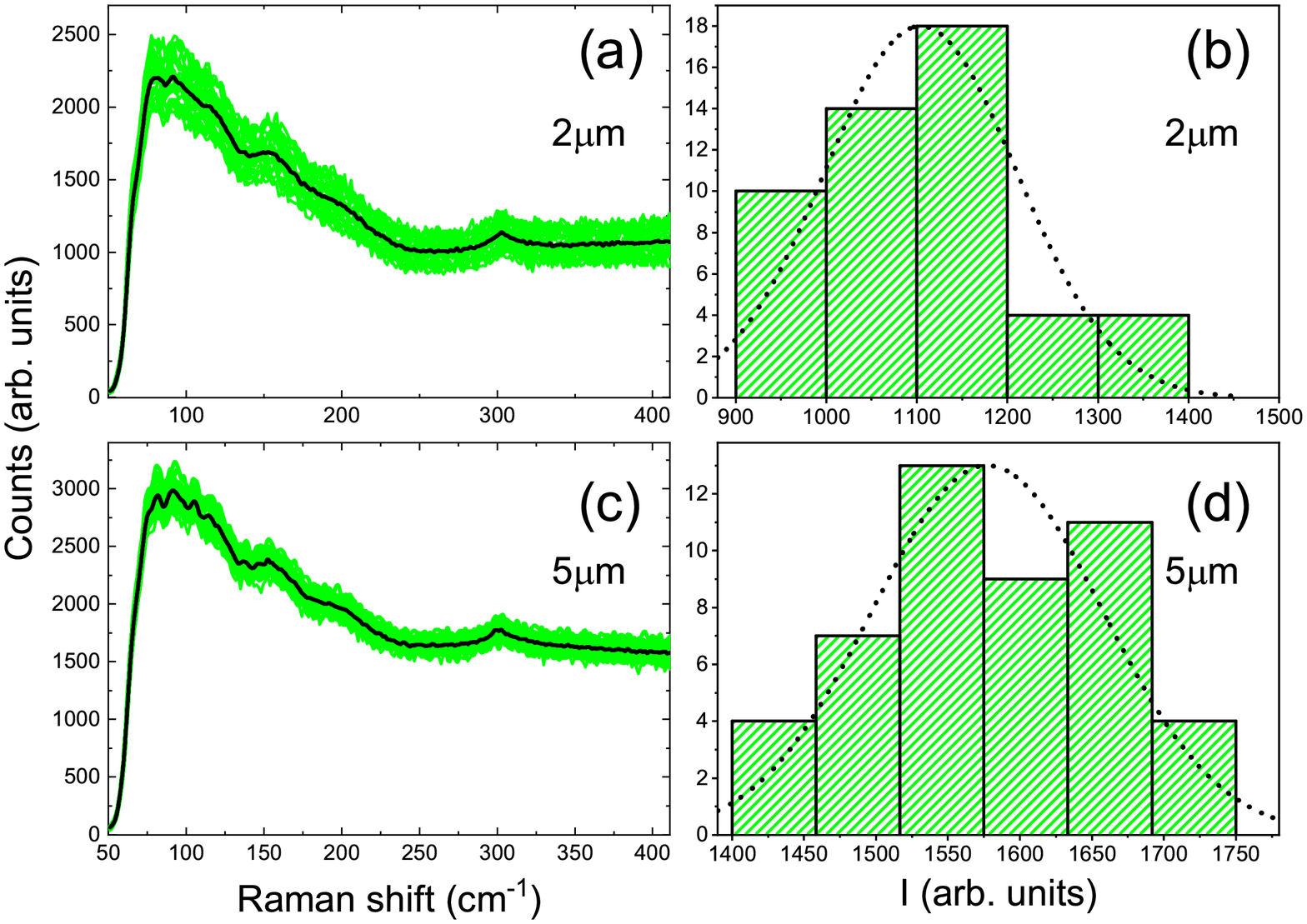}%
\caption{Raman spectra for the as-made In$_{\text{x}}$O sample with
\textit{N}$\approx$8x10$^{\text{20}}$cm$^{\text{-3}}.$ The sheet-resistance of
the film R$_{\square}$=18.34k$\Omega,$ at room-temperature which yields
k$_{\text{F}}\ell$=0.1. (a) and (c) show the Raman shift versus energy for
fifty different local sampling-points using 2$\mu$m and 5$\mu$m spot-size
respectively. The black line is the average of the 50 plots. (b) and (d)
exhibit the histogram of the Raman intensity-plots averaged over the interval
350-400 cm$^{\text{-1}}$ for each of the local readings related to the 2$\mu$m
and 5$\mu$m spot-size respectively. The histograms in (b) and (d) are fitted
to normal distributions depicted as dashed lines. These fits yield
heterogenous parameters (see text for definition) $\gamma$=9.7\% for the
2$\mu$m spot-size and $\gamma$=5.1\% for the 5$\mu$m spot-size.}%
\end{figure}
\begin{figure}[ptb]%
\centering
\includegraphics[
height=2.4829in,
width=3.4411in
]%
{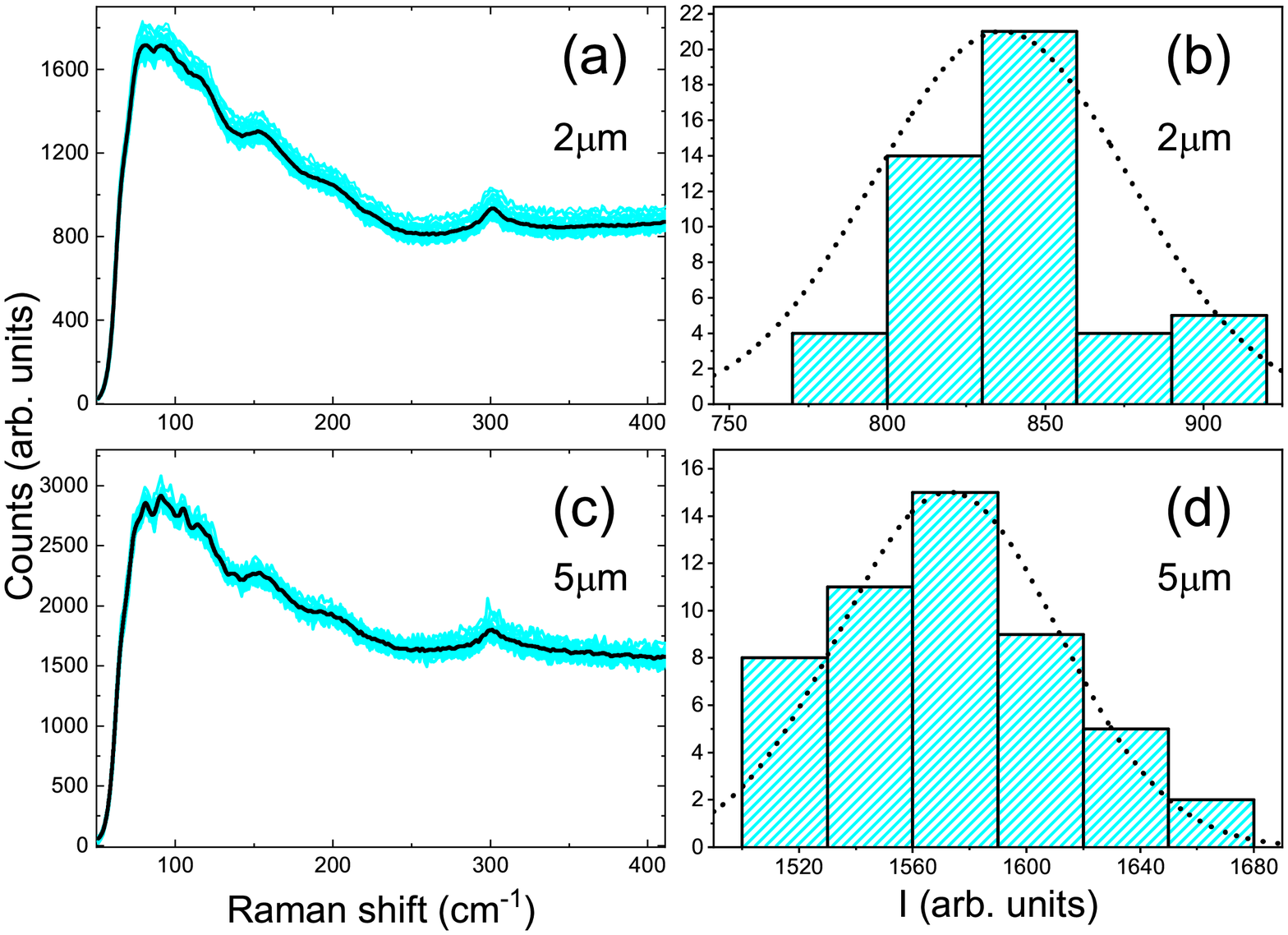}%
\caption{Raman spectra for the In$_{\text{x}}$O sample with \textit{N}%
$\approx$8x10$^{\text{20}}$cm$^{\text{-3}}$ after being heat-treated at
$\approx$380K for $\approx$48 hours. The sheet-resistance of the film changed
to: R$_{\square}$ =3.95k$\Omega$ which gives k$_{\text{F}}\ell$=0.42. (a) and
(c) show the Raman shift versus energy for fifty different local
sampling-points using 2$\mu$m and 5$\mu$m spot-size respectively. The black
line is the average of the 50 plots. (b) and (d) show the histogram of the
Raman intensity-plots averaged over the interval 350-400 cm$^{\text{-1}}$ for
each of the local readings related to the 2$\mu$m and 5$\mu$m spot-size
respectively. The histograms in (b) and (d) are fitted to normal distributions
depicted as dashed lines. These fits yield heterogenous parameters (see text
for definition) $\gamma$=4.9\% (down from $\gamma$=9.7\% in the as-made
sample) for the 2$\mu$m spot-size and $\gamma$=2.5\% (down from $\gamma$=5.1\%
in the as-made sample) for the 5$\mu$m spot-size.}%
\end{figure}

Before proceeding, the natural question is what is reflected by the magnitude
of the Raman signal?

At the energy range studied $\varepsilon$=70-420cm$^{\text{-1}}$, the signal
is composed of two components associated with different inelastic
light-scattering mechanisms. For energies $\varepsilon\leq$250cm$^{\text{-1}}%
$the main contribution comes from the boson peak \cite{26,27,28,29,30,31,32}
of the amorphous material. This feature, related to the phonon system of
In$_{\text{x}}$O, has been studied previously as function of disorder and
composition \cite{23}. The other component is presumably inelastic scattering
from the electronic states, either extended or weakly-localized in the
presence of quenched disorder.

To illustrate the relevance of this electronic mechanism for this material,
consider the Raman data shown in Fig. 3.
\begin{figure}[ptb]%
\centering
\includegraphics[
height=2.4785in,
width=3.4411in
]%
{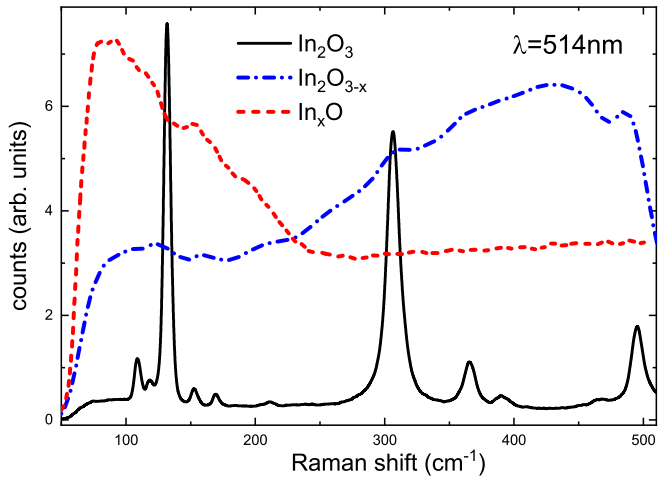}%
\caption{Raman spectra characteristic of three different versions of
indium-oxide films. The plot labeled as In$_{\text{2}}$O$_{\text{3}}$
(full-line) is based on data taken from the source material used in this work
for depositing the films. The In$_{\text{x}}$O (dashed line) is taken from a
90nm thick amorphous film. The polycrystalline version of indium-oxide
(dotted-dashed line) was crystallized and further heat-treated at $\approx$700
K for 25 min. to form In$_{\text{2}}$O$_{\text{3-x}}$ .The extended
heat-treatment at elevated temperature was needed to remove any residue of
amorphous component. The latter would otherwise show up as an excess signal at
the boson peak energy. Note that the boson peak does not appear in either of
the polycrystalline versions.}%
\end{figure}
Note the different spectra for the two polycrystalline versions of
indium-oxide. These samples share the same crystal structure with a
body-centered cubic symmetry \cite{33}. The grain-size of the In$_{\text{2}}%
$O$_{\text{3}}$ may be somewhat larger than in the deposited film
In$_{\text{2}}$O$_{\text{3-x}}$ but this cannot explain the grossly different
Raman spectra. A likely relevant difference between the compounds is their
chemical stoichiometry; the vacuum-deposited film In$_{\text{2}}%
$O$_{\text{3-x}}$ is 8-10\% oxygen deficient \cite{33} that, in turn, renders
the system metallic with carrier-concentration \textit{N}$\approx
$5x10$^{\text{19}}$cm$^{\text{-3}}$. The In$_{\text{2}}$O$_{\text{3}}$ sample
on the other hand is actually an intrinsic semiconductor with \textit{N}%
$\approx$10$^{\text{16}}$cm$^{\text{-3}}$ as measured by Hall effect at
room-temperature. It is then plausible to associate the distinctive spectrum
of the In$_{\text{2}}$O$_{\text{3-x}}$ with an electronic mechanism for
inelastic light-scattering \cite{34}, similar to the findings in underdoped
cuprate \cite{35}. 

Evidently, the electronic contribution to the Raman signal is substantial. It
overwhelms the Raman-active material vibration-modes and those that appear are
shifted in energy from the positions of their stoichiometric compound.

Note that the prominent boson peak portrayed by the amorphous sample is absent
(or is very weak) in either crystalline specimen. This is a demonstration of
the difference sensitivity of phonons and electrons to different types of
disorder; the grain-size in both, In$_{\text{2}}$O$_{\text{3-x}}$ films and
In$_{\text{2}}$O$_{\text{3}}$ sputtering-target pieces is at least 0.5$\mu$m,
much larger than the mean-free-path for electrons in samples where
k$_{\text{F}}\ell$%
$<$%
1. The large wavelength of phonons makes them less sensitive to the
short-range fluctuations of the chemical composition which strongly affects
charge carriers. Hence the effective Ioffe-Regel parameter for phonons
(k$_{\text{F}}\ell$)$_{\text{ph}}$ may be larger than unity, thus unfavorable
to exhibit a boson peak in the phonon density-of-states \cite{28}.

In contrast with good metals, Raman signal from `free' electrons is possible
in strongly disordered metals. The underlying physics has been considered in
\cite{34}. The theory strictly applies to k$_{\text{F}}\ell$%
$>$%
1 regime yet light-induced fluctuations of the chemical potential and the
ensuing relaxation of electron-hole pairs should still survive in
strongly-localized In$_{\text{x}}$O films. The intensity of low-frequency
radiation depends mainly on the charge-concentration as demonstrated in
high-T$_{\text{C}}$ compounds \cite{35}. This means that the fluctuation of
the Raman intensity depicted in Fig. 1 and Fig. 2 essentially signify the
variations of carrier-concentrations across these region of the sample. This
facet of the material affects directly and indirectly the disorder perceived
by charge-carriers. The width of the distribution, quantified by the value of
$\gamma$, is therefore a meaningful measure of the sample heterogeneity
averaged over a given length-scale.

It is remarkable that significant fluctuations of carrier-concentration are
observable over a scale of microns in the high-\textit{N} sample; $\gamma$ for
the 2$\mu$m area is $\approx$10\% in the as-made sample and is still $\approx
$5\% after being annealed (Fig. 2b). Moreover, it is evident from the
distribution (Fig. 1b and Fig. 2b) that the probability to find micron-size
areas with carrier-concentrations that differ by 20-30\% is appreciable. This
aspect has consequences for the low temperature transport in such systems in
particular, when the observed property is sensitive to the
carrier-concentration (for example; superconductivity). Indeed, high-\textit{N
}versions of In$_{\text{x}}$O exhibit anomalous transport effects near their
SIT that are ascribed to superconducting islands embedded in an insulating
matrix \cite{25} reflecting areas where \textit{N} is larger than the average
value of the sample. Similar emergence of disorder induced "granularity" was
found in other metal-oxides like high-T$_{\text{C}}$ compounds
\cite{36,37,38,39,40,41}.

It is illuminating to compare the before-and-after change in $\gamma$ for this
\textit{N}$\approx$8x10$^{\text{20}}$cm$^{\text{-3}}$ version of the material
(Fig. 1 and Fig. 2) with the respective behavior in the low-\textit{N} version
sample shown in Fig. 4 and Fig. 5 respectively.%

\begin{figure}[ptb]%
\centering
\includegraphics[
height=2.4595in,
width=3.4411in
]%
{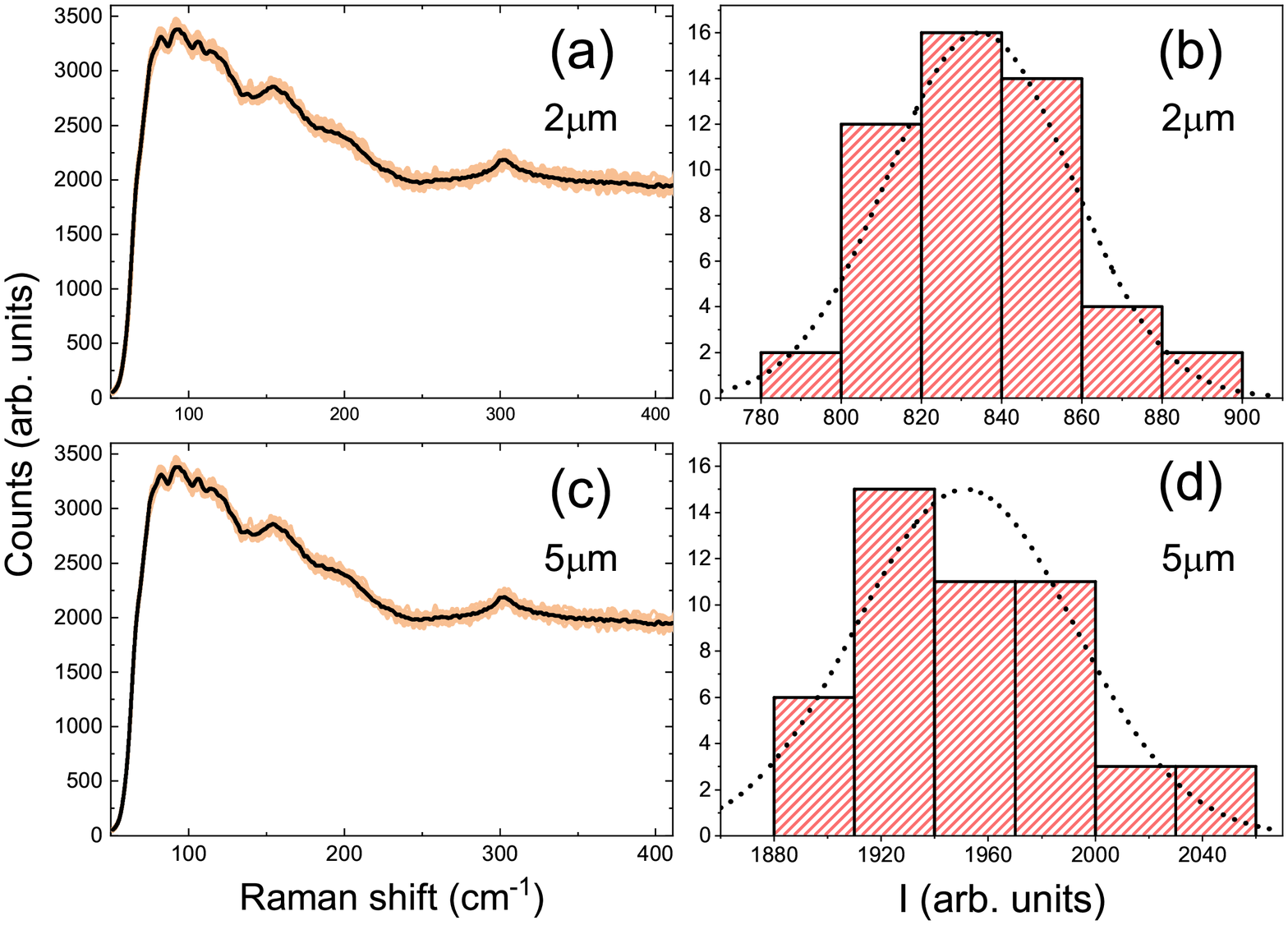}%
\caption{Raman spectra for the as-made In$_{\text{x}}$O sample with
\textit{N}$\approx$1x10$^{\text{19}}$cm$^{\text{-3}}.$ The original
sheet-resistance of the film R$_{\square}$=1.3M$\Omega,$ at room-temperature
which yields k$_{\text{F}}\ell$=0.005 (a) and (c) show the Raman shift versus
energy for fifty different local sampling-points using 2$\mu$m and 5$\mu$m
spot-size respectively. The black line is the average of the 50 plots. (b) and
(d) exhibit the histogram of the Raman intensity-plots averaged over the
interval 350-400 cm$^{\text{-1}}$ for each of the local readings related to
the 2$\mu$m and 5$\mu$m spot-size respectively. The histograms in (b) and (d)
are fitted to normal distributions depicted as dashed lines. These fits yield
heterogenous parameters (see text for definition) $\gamma$=2.7\% for the
2$\mu$m spot-size and $\gamma$=2.1\% for the 5$\mu$m spot-size.}%
\end{figure}
%

\begin{figure}[ptb]%
\centering
\includegraphics[
height=2.4379in,
width=3.4411in
]%
{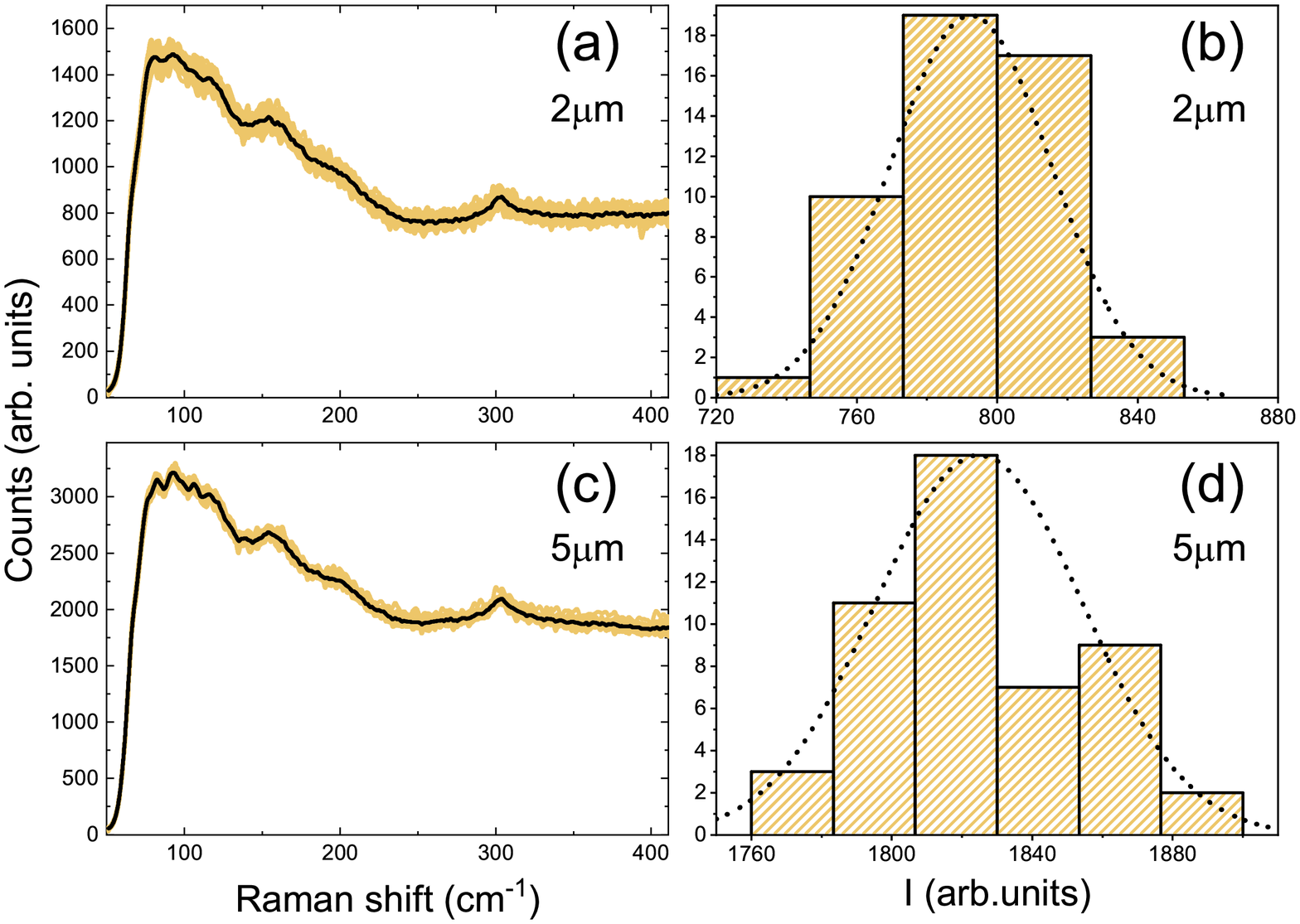}%
\caption{Raman spectra for the In$_{\text{x}}$O sample with \textit{N}%
$\approx$10$^{\text{19}}$cm$^{\text{-3}}$ after being heat-treated at
$\approx$380K for $\approx$48 hours. The sheet-resistance of the film changed
to :R$_{\square}$ =23k$\Omega$ which gives k$_{\text{F}}\ell$=0.29. (a) and
(c) show the Raman shift versus energy for fifty different local
sampling-points using 2$\mu$m and 5$\mu$m spot-size respectively. The black
line is the average of the 50 plots. (b) and (d) show the histogram of the
Raman intensity-plots averaged over the interval 350-400 cm$^{\text{-1}}$ for
each of the local readings related to the 2$\mu$m and 5$\mu$m spot-size
respectively. The histograms in (b) and (d) are fitted to normal distributions
depicted as dashed lines. These fits yield heterogenous parameters (see text
for definition) $\gamma$=2.8\% (compared with $\gamma$=2.7\% in the as-made
sample) for the 2$\mu$m spot-size and $\gamma$=1.6\% (down from $\gamma$=2.1\%
in the as-made sample) for the 5$\mu$m spot-size.}%
\end{figure}

Just by eyeing the raw data (compare Fig. 2 and Fig. 4) it is clear that even
the as-made low-\textit{N} version of In$_{\text{x}}$O is more homogenous than
the heat-treated specimen of the indium-rich material. It is gratifying to see
that the respective values of $\gamma$ are consistent with this observation.
This is in line with the notion that heterogeneity is an inherent aspect of
disorder in these In$_{\text{x}}$O films. Table 1 summarizes the values of
$\gamma$ for the respective samples and states.

Note that the change in $\gamma$ relative to the change in k$_{\text{F}}\ell$
is considerably weaker in the low-\textit{N} version.

To understand the reason for these observations and, in particular, the
steeper $\partial\gamma$/$\Delta$(k$_{\text{F}}\ell$) in the indium-rich
version, we need to expound on the connection between disorder and
k$_{\text{F}}\ell$.

\begin{center}%
\begin{tabular}
[c]{c|c|c|c|c|c|}\cline{2-6}
& \multicolumn{2}{|c|}{As-made $\gamma$(\%)} &
\multicolumn{2}{|c|}{Heat-treated $\gamma$(\%)} & $\Delta$(k$_{\text{F}}\ell
$)\\\hline
\multicolumn{1}{|c|}{Spot-size} & 2$\mu$m & 5$\mu$m & 2$\mu$m & 5$\mu$m &
\\\hline
\multicolumn{1}{|c|}{High-\textit{N}} & 9.7$\pm$0.5 & 5.1$\pm$0.5 & 4.9$\pm
$0.5 & 2.5$\pm$0.5 & 0.32\\\hline
\multicolumn{1}{|c|}{Low-\textit{N}} & 2.7$\pm$0.4 & 2.1$\pm$0.4 & 2.8$\pm
$0.4 & 1.6$\pm$0.4 & 0.29\\\hline
\end{tabular}

\end{center}

\begin{flushleft}
Table 1: The inhomogeneous parameter $\gamma$ (defined in the text), and the
change of the sample Ioffe-Regel parameter $\Delta$(k$_{\text{F}}\ell$) by
heat treatment for the two studied samples. The carrier concentration for the
high-\textit{N} and low\textit{-N }samples are \textit{N}=(8$\pm
$1)x10$^{\text{20}}$cm$^{\text{-3}}$ and \textit{N}=(1.5$\pm$1)x10$^{\text{19}%
}$cm$^{\text{-3}}$ respectively.
\end{flushleft}

In the context of transport, one usually weighs disorder on an \textit{energy}
scale while k$_{\text{F}}\ell$ is \textit{dimensionless}. A quantitative
estimate of disorder is therefore limited to special values of k$_{\text{F}%
}\ell$ such as its value at the Anderson MIT where the disorder-energy may be
assessed with respect to E$_{\text{F}}$ \cite{42}. This means that, without
specific information on auxiliary properties of a given system, k$_{\text{F}%
}\ell$ is merely a relative measure of disorder.

Moreover, k$_{\text{F}}\ell$ may be relied on to be indicative of
quenched-disorder when $\rho_{\text{RT}}$ is strictly determined by elastic
scattering, a condition that may be obeyed for k$_{\text{F}}\ell$ values at
the vicinity of the MIT. The critical k$_{\text{F}}\ell$ in In$_{\text{x}}$O
occurs at k$_{\text{F}}\ell$=0.31$\pm$0.2 essentially independent of the
carrier-concentration \cite{25} presumably because the disorder in these
samples is always larger than the interaction \cite{43,44}. On the other hand,
versions of this material with different \textit{N} have different\textit{
}degrees of disorder for the same value of\textit{ }k$_{\text{F}}\ell$. In
particular, the disorder that brings the system near the Anderson transition,
is greater the larger \textit{N} is. Recall that the critical disorder is
associated with potential-fluctuations that are large enough to overcome the
Fermi-energy E$_{\text{F}}\,$\cite{42} that in turn increases with the
carrier-concentration of the system.

For the same reason, for higher \textit{N}, a larger degree of disorder is
required to achieve a given change in k$_{\text{F}}\ell$. This may be
illustrated by observing the change in the material optical properties
accompanying a reduction of resistivity following heat-treatment. Figure 6
shows how the optical-gap E$_{\text{g}}$ changes after several heat-treatments
\cite{45} that progressively increase k$_{\text{F}}\ell$ by reducing
$\rho_{\text{RT}}$. E$_{\text{g}}$ signifies the energy scale for transitions
between the valence and the conduction bands of In$_{\text{x}}$O. The figure
compares the E$_{\text{g}}$ vs. k$_{\text{F}}\ell$ dependence for two versions
of In$_{\text{x}}$O with different carrier-concentrations. A change of
E$_{\text{g}}$ is a consequence of the system densification \cite{24}.
Rearrangement of oxygen vacancies take place concomitantly. Both processes
reduce the system disorder. Figure 6 shows that a larger degree of structural
change, reflected by $\Delta$E$_{\text{g}}$, is required to induce a similar
change in k$_{\text{F}}\ell$ for the high-\textit{N} version of In$_{\text{x}%
}$O. This is an example of a relation between the energy change required to
affect $\Delta$(k$_{\text{F}}\ell$) in a specific system.%

\begin{figure}[ptb]%
\centering
\includegraphics[
height=2.373in,
width=3.4411in
]%
{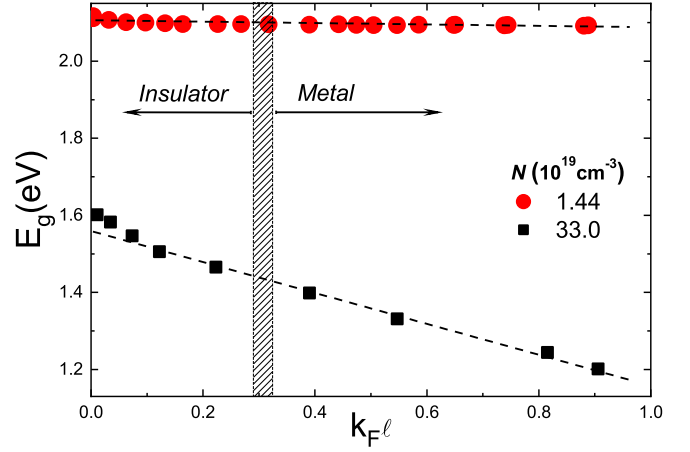}%
\caption{The dependence of the optical-gap E$_{\text{g}}$ on the Ioffe-Regel
k$_{\text{F}}\ell$ parameter for two batches of In$_{\text{x}}$O samples
labeled by their respective \textit{N}. Dashed lines are guides to the eye.
Note the much steeper dependence of E$_{\text{g}}$ versus a change in
k$_{\text{F}}\ell$ for the sample with the larger \textit{N}.}%
\end{figure}

The two versions of the material differ by their chemical composition,
explicitly, by their O/In ratio. For the low-\textit{N} version in Fig. 6 the
ratio is $\approx$1.4 as compared with 1.5 for the stable compound
In$_{\text{2}}$O$_{\text{3}}$ while it is $\approx$1.25 for the indium-rich
version \cite{25}. This relatively small difference in composition leads to a
substantial difference in carrier-concentration; the high-\textit{N} has a
larger carrier-concentration than the low-\textit{N} sample in Fig. 6 by a
factor of $\approx$23. This means that their Fermi-energies differ by a factor
of\textit{ }$\approx$9 which accounts for the difference in the slopes for
$\Delta$E$_{\text{g}}$/$\Delta$(k$_{\text{F}}\ell$) of the samples. The higher
sensitivity of the conductance of the low-\textit{N} system to changes in the
ions-potential was offered \cite{46} as the physical reason for the
observation that the magnitude of the 1/f-noise often scales with the inverse
of the carrier-concentration (the "Hooge law" \cite{47}).%

\begin{figure}[ptb]%
\centering
\includegraphics[
height=4.6311in,
width=3.4411in
]%
{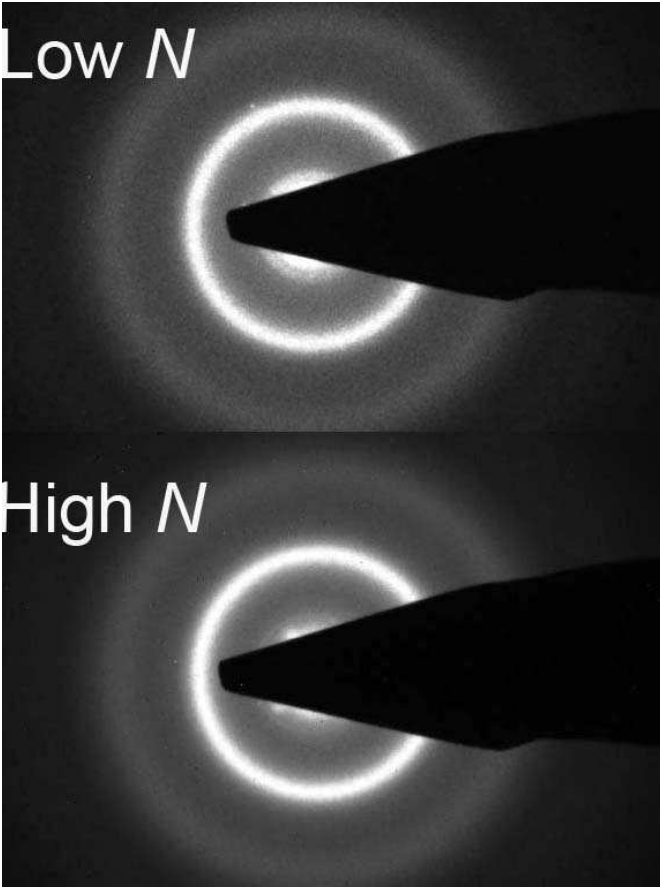}%
\caption{Electron diffraction patterns for two In$_{\text{x}}$O films
deposited on carbon-coated Cu grids. Both films have identical thicknesses of
d$\approx$20 nm (determined by the quartz-crystal monitor). The top pattern,
labeled Low-\textit{N}, has \textit{N}$\approx$1x10$^{\text{19}}%
$cm$^{\text{-3}}$ while the High-\textit{N} sample has \textit{N}$\approx
$9x10$^{\text{20}}$cm$^{\text{-3}}$. The patterns appear quite similar but a
careful comparison of intensity shows an overall higher signal in the
indium-rich film. This is due to that the dominant signal is determined by the
indium atoms.}%
\end{figure}

The fact that, for a similar k$_{\text{F}}\ell$, the high-\textit{N} sample is
more disordered than the low-\textit{N} sample is not evident from the
diffraction patterns of the samples. From the point of view of topology, these
two versions are difficult to tell apart; the diffraction patterns comparing
between low-\textit{N} and high-\textit{N} versions of In$_{\text{x}}$O films
shown in Fig. 7 reveal amorphous structures with nearly identical
nearest-neighbor distance for both versions (judging be the strong
diffraction-ring diameter).

The additional source of disorder that accounts for the difference between the
two versions in Fig. 7 is their deviation from the chemical stoichiometry of
the compound. This property is carried by the relatively few valence electrons
while diffraction patterns are dominated by scattering from the plentiful
inner-shell electrons. Charge-carriers that are involved in transport, on the
other hand, are equally sensitive to both types of disorder.

To understand the origin of the disorder associated with the lack of
stoichiometry, recall that the value of \textit{N} is determined by the O/In
ratio. Oxygen deficiency in indium-oxide specimens is of the order of 10-20\%
while the observed \textit{N} is typically smaller by 2-3 orders of magnitude.
Therefore some indium atoms must have a valency of +1 rather than the more
common +3 to account for the observed \textit{N} \textit{and} preserve
chemical neutrality \cite{33}.The relative number of these ions determines the
carrier-concentration while the way they are arranged in space affects the
disorder perceived by charge-carriers.

The scattering centers for the conduction electrons are actually oxygen
vacancies (and possibly, di-vacancies). Being the lighter element in the
compound, oxygen atoms are also the dominant species to diffuse through the
sample during heat-treatment. Measurement of carrier-concentration showed it
to be essentially constant during treatment and therefore oxygen movement is
essentially confined to within the boundaries of the sample. The constancy of
the global carrier-concentration was confirmed by three independent means;
Hall effect \cite{12}, optical absorption \cite{42}, and by monitoring the
position of the boson-peak \cite{23}. Therefore, migration of oxygen is
plausibly the underlying mechanism for the changes in the spatial
rearrangement of oxygen vacancies that take place during heat-treatment.

To be consistent with our observations, the oxygen diffusion-coefficient
D$_{\text{O}}$ should be no smaller than $\approx$10$^{\text{-12}}$
cm$^{\text{2}}$/s; this will make it feasible for oxygen to diffuse over a
region L$\approx$5$\mu$m during a time $\tau\approx$two-days, the duration of
heat-treatment in the present study. This value is comparable with
conventionally measured D$_{\text{O}}$ for oxygen diffusion in several solids
\cite{48} and in a polymer glass \cite{49} (extrapolating results of these
studies to the typical temperature used in the heat treatment, T$\approx
$360K). Actually, the presence of vacancies probably makes oxygen diffusion in
these In$_{\text{x}}$O \ thin films faster than in the bulk solids we compare
it with.

The dynamics of processes that occur on a microscopic scales depends on
multi-stage bond-formation and local structure rearrangement \cite{24,50} that
probably involves many-particle coordinated transitions.

In addition to the sample conductance, a feature that is affected by these
short-range processes is the film surface roughness. This aspect may be probed
by X-ray interferometry (XRR) using typically a sub-nanometer wavelength
radiation. An example of XRR spectra taken before and after a brief
heat-treatment of a high-\textit{N} In$_{\text{x}}$O film is shown in Fig. 8.%

\begin{figure}[ptb]%
\centering
\includegraphics[
height=2.5391in,
width=3.4411in
]%
{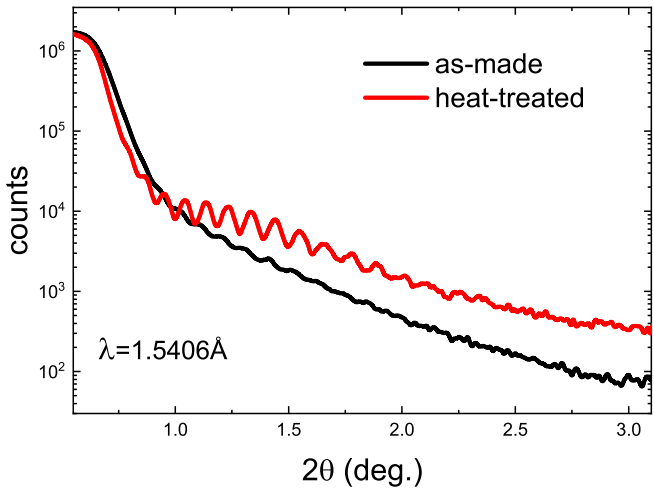}%
\caption{X-ray reflectometry spectra for a high-\textit{N} sample
(\textit{N}$\approx$8.5x10$^{\text{20}}$cm$^{\text{-3}}$) version of
In$_{\text{x}}$O film having similar composition as that of the sample in Fig.
1 but with a thickness d=71nm. Plots are shown for the before-and-after
heat-treatment at 360$\pm$5K for 14 hours. This resulted in the
sheet-resistance of the film to decrease from R$_{\square}$=6.8k$\Omega$ to
R$_{\square}$=3.12k$\Omega$. Curves are displaced along the ordinate for
clarity.}%
\end{figure}

The enhanced visibility of the XRR interference pattern following
heat-treatment is the result of a more homogeneous distribution of defects.
These processes take place on a microscopic scale, directly affecting the
conductivity via changes in the basic transport parameters such as
mean-free-path, or the hopping-length in the diffusive or hopping regime respectively.

The effect on conductivity due to homogenization of the \textit{mesoscopic}
regions probed by the Raman scattering is a more intricate issue. The length
scale relevant for conductivity in In$_{\text{x}}$O films are much smaller
than a few microns whether the system is on the metallic or insulating side of
the transition even at cryogenic temperatures.

As mentioned above, the macroscopic carrier-concentration of the sample is
essentially unchanged during heat-treatment. Local changes in \textit{N}, on
the other hand, do occur as evidenced by the Raman data (Fig. 1 and Fig. 2).
These changes may increase the conductance of a given region while decreasing
the conductance of another region (to be consistent with the global condition
on \textit{N}). The change in the macroscopic conductance due to these
homogenization events will therefore be small except deep in the insulating
side where the sample conductance would increase due to the logarithmic
averaging inherent to the hopping regime \cite{51,52}.

In summary, we examined in this work the relation between disorder and spatial
uniformity of the potential perceived by the charge carriers in an amorphous
metal. Films of In$_{\text{x}}$O with different carrier-concentration but
similar k$_{\text{F}}\ell$ were used to represent systems that differ
significantly by their absolute degree of disorder. Both systems exhibit the
same `frozen-liquid' structure with essentially indistinguishable diffraction
patterns but they differ in their `defects' density determined by their
chemical composition. It is shown that the film with the stronger disorder
exhibits a higher degree of heterogeneity extending over mesoscopic scales.
Moreover, reducing the disorder in the film by heat-treating it, resulted in a
decrease in heterogeneity over both microscopic, and mesoscopic scales. In
this case the change of disorder that resulted from the treatment is
characterized by the Ioffe-Regel parameter acting as a relative measure of
disorder. Disorder in In$_{\text{x}}$O is presumably dominated by the local
concentration of oxygen vacancies that play the dual role as dopants and
scatterers. Charge carriers that originate from off-stoichiometry conditions
is a common scenario in many oxygen-deficient metal-oxides including
high-T$_{\text{C}}$ compounds. It would be interesting to see how general is
the relation between disorder and heterogeneity in these compounds and in
other disordered materials.

Another corollary implicit to the arguments raised above is that local
defects, the oxygen vacancies in the current system, contribute to
$\rho_{\text{RT}}$ more than the lack of long-range-order associated with
amorphicity. This means that, for the same k$_{\text{F}}\ell$, the larger is
the In/O ratio, the larger is the contribution of the diagonal component to
the system disorder. This makes In$_{\text{x}}$O a useful platform to study
experimentally the relative role of the two types of disorder.

\begin{acknowledgments}
The assistance by Dr. Anna Radko with the Raman spectra work is gratefully
acknowledged. This research has been partially supported by the 1030/16 grant
administered by the Israel Academy for Sciences and Humanities.
\end{acknowledgments}


\begin{thebibliography}{99}                                                                                               %


\bibitem {1}Bernard Derrida, Can disorder induce several phase transitions?
Physics Reports,\textbf{103}, 29 (1984).

\bibitem {2}M. Inui, S. A. Trugman, and Elihu Abrahams, Unusual properties of
midband states in systems with off-diagonal disorder, Phys. Rev. B.
\textbf{49}, 3190 (1994).

\bibitem {3}M. A. Ramos and S. Vieira, F. J. Bermejo and J. Dawidowski, H. E.
Fischer, H. Schober, M. A. Gonz\'{a}lez, C. K. Loong and D. L. Price,
Quantitative Assessment of the Effects of Orientational and Positional
Disorder on Glassy Dynamics, Phys. Rev. Lett., \textbf{78}, 82 (1997).

\bibitem {4}T. Bellini, M. Buscaglia, C. Chiccoli, F. Mantegazza,. P. Pasini,
and C. Zannoni, Nematics with Quenched Disorder: What Is Left when Long Range
Order Is Disrupted? Phys. Rev. Lett., \textbf{85}, 1008\ (2000).

\bibitem {5}S\`{e}bastien Balibar and Fr\'{e}d\'{e}ric Caupin, Supersolidity
and disorder, Journal of Physics: Condensed Matter, \textbf{20}, 173201 (2008).

\bibitem {6}Paraj Titum, Netanel H. Lindner, Mikael C. Rechtsman, and Gil
Refael, Disorder-Induced Floquet Topological Insulators, Phys. Rev. Lett.,
\textbf{114}, 056801 (2015).

\bibitem {7}Sa\u{s}o Grozdanov, Andrew Lucas,\ Subir Sachdev,\ and Koenraad
Schalm, Absence of Disorder-Driven Metal-Insulator Transitions in Simple
Holographic Models, Phys. Rev. Lett., \textbf{115}, 221601 (2015).

\bibitem {8}F. Baboux, L. Ge, T. Jacqmin, M. Biondi, E. Galopin, A.
Lema\^{\i}tre, L. Le Gratiet, I. Sagnes, S. Schmidt, H.\thinspace E.
T\"{u}reci, A. Amo, and J. Bloch, Bosonic Condensation and Disorder-Induced
Localization in a Flat Band, Phys. Rev. Lett., \textbf{116}, 066402 (2016).

\bibitem {9}Thomas Vojta, Disorder in Quantum Many-Body Systems, Annual Review
of Condensed Matter Physics, \textbf{10}, 233 (2019).

\bibitem {10}M. E. Raikh and E. V. Tsiper, Energy spectrum and size
quantization in partially ordered semiconductor alloys, Phys. Rev.B,
\textbf{49}, 2509\ (1994).

\bibitem {11}V. M. Apalkov and M. E. Raikh, Universal fluctuations of the
random lasing threshold in a sample of a finite area, Phys. Rev.B \textbf{71},
054203 (2005); T. A. Sedrakyan, E. G. Mishchenko, and M. E. Raikh, Zero-Bias
Tunneling Anomaly in a Clean 2D Electron Gas Caused by Smooth Density
Variations, Phys. Rev. Lett., \textbf{99}, 206405 (2007).

\bibitem {12}Z. Ovadyahu, Some finite temperature aspects of the Anderson
transition, J. Phys. C: Solid State Phys., \textbf{19}, 5187 (1986).

\bibitem {13}D. Bruce Buchholz, Qing Ma, Diego Alducin, Arturo Ponce, Miguel
Jose-Yacaman, Rabi Khanal, Julia E. Medvedeva, and Robert P. H. Chang, The
Structure and Properties of Amorphous Indium Oxide, Chem. Mater. \textbf{26},
5401 (2014).

\bibitem {14}D. Shahar and Z. Ovadyahu, Superconductivity near the Mobility
Edge, Phys. Rev. B \textbf{46}, 10917 (1992).

\bibitem {15}V. Gantmakher, Transport properties of normal and quasinormal
states of poor superconductors,International Journal of Modern Physics B,
\textbf{12}, 29 (1998).

\bibitem {16}D. Kowal and Z. Ovadyahu, Disorder induced granularity in an
amorphous superconductor, Solid State Comm., \textbf{90}, 783 (1994).

\bibitem {17}G. Sambandamurthy, L. W. Engel, A. Johansson, and D. Shahar,
Superconductivity-related insulating behavior, Phys. Rev. Lett. \textbf{92},
107005 (2004).

\bibitem {18}S. Poran, E. Shimshoni, and A. Frydman, Disorder-induced
superconducting ratchet effect in nanowires, jPhys. Rev. B \textbf{84}, 014529 (2011).

\bibitem {19}Benjamin Sac\'{e}p\'{e}, Thomas Dubouchet, Claude Chapelier, Marc
Sanquer, Maoz Ovadia, Dan Shahar, Mikhail Feigel'man and Lev Ioffe,
Localization of preformed Cooper pairs in disordered superconductors, Nature
Physics, \textbf{7}, 239 (2011).

\bibitem {20}D. Sherman, G. Kopnov, D. Shahar, and A. Frydman, Phys. Rev.
Lett. \textbf{108}, 177006 (2012).

\bibitem {21}Yeonbae Lee, Aviad Frydman, Tianran Chen, Brian Skinner, and A.
M. Goldman, Electrostatic tuning of the properties of disordered indium-oxide
films near the superconductor-insulator transition, Phys. Rev. B \textbf{88},
024509 (2013).

\bibitem {22}Daniel Sherman, Uwe S. Pracht, Boris Gorshunov, Shachaf Poran,
John Jesudasan, Madhavi Chand, Pratap Raychaudhuri, Mason Swanson, Nandini
Trivedi, Assa Auerbach, Marc Scheffler, Aviad Frydman \& Martin Dresse, The
Higgs mode in disordered superconductors close to a quantum phase transition,
Nature Physics volume \textbf{11}, 188 (2015).

\bibitem {23}Itai Zbeda, Ilana Bar and Z. Ovadyahu, Microstructure and the
boson peak in thermally treated In$_{\text{x}}$O films, Phys. Rev. Mat.
\textbf{5}, 085602 (2021).

\bibitem {24}Z. Ovadyahu, Memory versus irreversibility in the thermal
densification of amorphous glasses, Phys. Rev. B \textbf{95}, 214207 (2017);
\textit{ibid}, Structure dynamics in thermal-treatment of amorphous
indium-oxide films, Phys. Status Solidi B \textbf{257},1900310 (2020).

\bibitem {25}U. Givan and Z. Ovadyahu, Compositional disorder and transport
peculiarities in the amorphous indium-oxides, Phys. Rev. B \textbf{86}, 165101 (2012).

\bibitem {26}S. R. Elliott, A Unified Model for the Low-Energy Vibrational
Behavior of Amorphous Solids, EPL \textbf{19} 201 (1992)

\bibitem {27}Walter Schirmacher, Gregor Diezemann, and Carl Ganter, Harmonic
vibrational excitations in disordered solids and the \textquotedblleft boson
peak\textquotedblright, Phys. Rev. Lett. \textbf{81}, 136 (1998).

\bibitem {28}H. Shintani and H. Tanaka, Universal link between the boson peak
and transverse phonons in glass, Nature Mater. \textbf{7}, 870 (2008).

\bibitem {29}V. L. Gurevich, D. A. Parshin, and H. R. Schober, Pressure
dependence of the boson peak in glasses, Phys. Rev. B \textbf{71}, 014209 (2005).

\bibitem {30}H. R. Schober, U. Buchenau, and V. L. Gurevich, Pressure
dependence of the boson peak in glasses: Correlated and uncorrelated
perturbations, Phys. Rev. B \textbf{89}, 014204 (2014).

\bibitem {31}V. K. Malinovsky and A. P. Sokolov, The neature of boson peak in
Raman scattering in glasses, Solid State Comm., \textbf{57}, 757(1986).

\bibitem {32}Walter Schirmacher, Tullio Scopigno, Giancarlo Ruocco, Theory of
vibrational anomalies in glasses, Journal of Non-Crystalline Solids
\textbf{407}, 133 (2015).

\bibitem {33}Z. Ovadyahu, B. Ovryn and H.W. Kraner, Microstructure and
electro-optical properties of evaporated indium-oxide Films, J. Elect. Chem.
Soc. \textbf{130}, 917 (1983).

\bibitem {34}A. Zawadowski and M. Cardona,Theory of Raman scattering on normal
metals with impurities, Phys. Rev.\textbf{42}, 10 732 (1990).

\bibitem {35}R. Nemetschek, M. Opel, C. Hoffmann, P. F. M\"{u}ller, R. Hackl,
H. Berger and L. Forr\'{o}, A. Erb and E. Walker, Pseudogap and
Superconducting Gap in the Electronic Raman Spectra of Underdoped Cuprates,
Phys. Rev. Lett. \textbf{78}, 4837 (1997).

\bibitem {36}J. A. Hoyos, C. Kotabage, and T. Vojta, Effects of dissipation on
a quantum critical point with disorder, Phys. Rev. Lett. \textbf{99}, 230601 (2007).

\bibitem {37}S. Ubaid-Kassis, T. Vojta, and A. Schroeder, Quantum Griffiths
phase in the weak itinerant ferromagnetic alloy Ni$_{\text{1-x}}$V$_{\text{x}%
}$, Phys. Rev. Lett. \textbf{104}, 066402 (2010).

\bibitem {38}R. Wang, A. Gebretsadik, S. Ubaid-Kassis, A. Schroeder, T. Vojta,
P. J. Baker, F. L. Pratt, S. J. Blundell, T. Lancaster, I. Franke, J. S.
M\"{o}ller, and K. Page, Quantum Griffiths phase inside the ferromagnetic
phase of Ni$_{\text{1-x}}$V$_{\text{x}}$, Phys. Rev. Lett. \textbf{118},
267202 (2017).

\bibitem {39}Y. Xing, K. Zhao, P. Shan, F. Zheng, Y. Zhang, H. Fu, Y. Liu, M.
Tian, C. Xi, H. Liu, J. Feng, X. Lin, S. Ji, X. Chen, Q.-K. Xue, and J.Wang,
Ising superconductivity and quantum phase transition in macro size monolayer
NbSe$_{\text{2}}$, Nano Lett. \textbf{17}, 6802 (2017).

\bibitem {40}Y. Saito, T. Nojima, and Y. Iwasa, Quantum phase transitions in
highly crystalline two-dimensional superconductors, Nature communications,
\textbf{9}, 1 (2018).

\bibitem {41}N. A. Lewellyn, I. M. Percher, J. Nelson, J. Garcia-Barriocanal,
I. Volotsenko, A. Frydman, T. Vojta, and A. M. Goldman, Infinite-randomness
fixed point of the quantum superconductor-metal transitions in amorphous thin
films, Phys. Rev. B \textbf{99}, 054515 (2019).P. Reiss, D. Graf, A. A.
Haghighirad, T. Vojta, and A. I. Coldea, Signatures of a quantum Griffiths
phase close to an electronic nematic quantum phase transition, Phys. Rev.
Lett. \textbf{127}, 246402 (2021).

\bibitem {42}Z. Ovadyahu, Slow dynamics of the electron-glasses; the role of
disorder, Phys. Rev. B. \textbf{95}, 134203 (2017).

\bibitem {43}E. Yamaguchi, H. Aoki and H. Kamimura, Intra- and interstate
interactions in Anderson localised states, J. Phys. C: Solid State Phys.,
\textbf{12}, 4801 (1979); H. Kamimura, Theoretical model on the interplay of
disorder and electron correlations, Progress of Theoretical Physics
Supplement, \textbf{72}, 206 (1982).

\bibitem {44}Z. Ovadyahu, Interaction-induced spatial correlations in a
disordered glass, Phys. Rev. B \textbf{105}, 235101 (2022).

\bibitem {45}The samples in figure 6 are part of a fuller study described in
\cite{42}.

\bibitem {46}O. Cohen and Z. Ovadyahu, .Resistance Noise near the Anderson
Transition, Phys. Rev. B \textbf{50}, 10442 (1994).

\bibitem {47}F. N. Hooge, 1/f noise is no surface effect, Phys. Lett.
\textbf{29A}, 139 (1969).

\bibitem {48}R. Kirchheim, Metals as sinks and barriers for interstitial
diffusion with examples for oxygen diffusion in copper, niobium, and tantalum,
Acta Metallurgica, \textbf{27}, 869 (1979).

\bibitem {49}Lars Poulsen and Peter R. Ogilby, Oxygen Diffusion in Glassy
Polymer Films: Effects of Other Gases and Changes in Pressure, J. Phys. Chem.
\textbf{104}, 2573 (2000).

\bibitem {50}Yu. M. Galperin and V. I.. Gurevich, Theory of low-temperature
thermal expansion of glasses, 32, 6873 (1985).

\bibitem {51}Hui Lin Zhao, Boris Z. Spivak, Martin P. Gelfand, and Shechao
Feng, Negative magnetoresistance in variable-range-hopping conduction, Phys.
Rev. B \textbf{44}, 10 760 (1991).

\bibitem {52}O. Entin-Wohlman, Y. Imry, and U. Sivan, Orbital
magnetoconductance in the variable-range-hopping regime, Phys. Rev. B
\textbf{40}, 8342 (1989).
\end{thebibliography}
\end{document}